\documentclass[aps,pre,preprint,onecolumn,citeautoscript,superscriptaddress,eqsecnum]{revtex4-1}  
\usepackage{epsfig}
\usepackage{graphicx}
\usepackage{dcolumn}
\usepackage{caption}
\usepackage{subcaption}
\usepackage[bookmarks=false]{hyperref}
\hypersetup{colorlinks=true,
           citecolor=red,
            filecolor=blue,
           urlcolor=blue}
\usepackage{ulem}
\usepackage{bm}
\usepackage{color}
\usepackage{braket}
\usepackage{latexsym}
\usepackage{amsmath}
\usepackage{amssymb}
\usepackage{latexsym}

\def \beq{\begin{eqnarray}}
\def \eeq{\end{eqnarray}}
\def \r{{\mathbf{r}}}
\def \rp{{\mathbf{r}^{\prime}}}
\def \Q{{\mathbf{Q}}}
\def \tQ{{\widetilde{\mathbf{Q}}}}
\def \K{{\mathbf{K}}}

\def \k{{\mathbf{k}}}

\def \nn{\nonumber \\}

\begin{document}

\title{A fractionalized Fermi liquid with bosonic chargons\\ as a candidate for the pseudogap metal}

\author{Shubhayu Chatterjee}
\affiliation{Department of Physics, Harvard University, Cambridge Massachusetts
02138, USA.}
\author{Subir Sachdev}
\affiliation{Department of Physics, Harvard University, Cambridge Massachusetts
02138, USA.}
\affiliation{Perimeter Institute of Theoretical Physics, Waterloo Ontario-N2L 2Y5, Canada.}

\date{\today \\
\vspace{1.6in}}
\begin{abstract}
Doping a Mott-insulating $\mathbb{Z}_2$ spin liquid can lead to a fractionalized Fermi liquid (FL*). Such a phase has several favorable features that make it a candidate for the pseudogap metal for the underdoped cuprates. We focus on a particular, simple
$\mathbb{Z}_2$-FL* state which
can undergo a confinement transition to a spatially uniform superconductor which is smoothly connected
to the `plain vanilla' BCS superconductor with $d$-wave pairing. Such a transition occurs by the condensation of bosonic particles carrying 
$+e$ charge but no spin (`chargons'). We show that modifying the dispersion of the bosonic chargons can lead to confinement transitions with charge density waves and pair density waves at the same wave-vector $\K$, co-existing with $d$-wave superconductivity. We also compute the evolution of the Hall number in the normal state during the transition from the plain vanilla FL* state to a Fermi liquid, and
argue, following Coleman, Marston, and Schofield \href{http://link.aps.org/doi/10.1103/PhysRevB.72.245111}{[Phys. Rev. B \textbf{72}, 245111 (2005)]}, that it exhibits a discontinuous jump near optimal doping. We note the distinction between these results and those
obtained from models of the pseudogap with fermionic chargons.
\end{abstract}
\maketitle
\tableofcontents

\section{Introduction}
Recent experiments on the hole-doped cuprates have demonstrated that the pseudogap (PG) phase behaves remarkably like a Fermi liquid. For example, both the temperature and frequency dependence of the optical conductivity $\left[ \sigma(\omega) \sim 1/(- i \omega + \tau^{-1}) \text{ with } \tau^{-1} \sim \omega^2 + T^2 \right]$ \cite{MirzaeiPNAS13}, as well as the consistency of magnetoresistance with Kohler's rule ($\rho_{xx} \sim \tau^{-1}[1 + (H\tau)^2]$) \cite{ChanPRL14}, are behavior typical of Fermi liquids. However, more 
recent measurements of Hall coefficient at high magnetic fields and low $T$ \cite{LTCP15,Laliberte2016} provides evidence for a crucial difference of this phase from a conventional Fermi liquid (FL). Doping a half-filled Mott insulator with a density of $p$ holes should lead to a hole-like Fermi surface of size $1+p$. Although this is indeed seen for large doping, the situation is different in the PG regime. In this regime, when additional Fermi surface reconstruction due to density waves are absent, the Hall coefficient corresponds to a Fermi surface of size $p$, which violates Luttinger's theorem \cite{LuttingerTh1960}. In absence of any symmetry breaking long-range order, this can be possible only in the presence of excitations of emergent gauge fields. A phase which realizes such a Fermi surface is called a fractionalized Fermi liquid (FL*) \cite{FFL,TSMVSS04,Ribeiro05,Ribeiro06,MeiWen12,MPSS12,Punk15}. 

The $\mathbb{Z}_2$-FL* is a viable candidate for the PG metal, as several properties of the phase can be understood from this point of view. First, the presence of the emergent $\mathbb{Z}_2$ gauge field allows it to violate Luttinger's theorem \cite{FFL,APAV04} without any long range symmetry-breaking order. Model calculations \cite{YQSS10} yield hole-pockets centered near $(\pm \pi/2, \pm \pi/2)$ with an anisotropic electron quasiparticle residue, which can explain the observation of Fermi arcs in photoemission experiments \cite{DamascelliRMP2003,VishikShen2010}. Further, density wave instabilities of the $\mathbb{Z}_2$-FL* naturally lead to $d$-form factor bond density wave with charge modulation on the bonds and a wave-vector similar to STM observations \cite{DCSS14,hamidian_2015nphys}. Such density waves were also shown to arise via a different route from a $\mathbb{Z}_2$-FL*, through a confinement transition that destroys topological order \cite{PCAS16}. Superconductivity also appears naturally as a descendant of a $\mathbb{Z}_2$-FL*, as pairing between emergent fractionalized excitations spinons are an inherent characteristic of such a phase, and this can mediate pairing between the electron-like quasiparticles which form the small Fermi surface \cite{FFL}. 

While the above are rather general properties of the $\mathbb{Z}_2$-FL* state, more thorough considerations lead
to significant observable differences between different realizations of such a state. In particular, it is useful to distinguish
between $\mathbb{Z}_2$-FL* states in which the lowest energy excitations which carry charge but no spin (``chargons")
are fermionic or bosonic. Models with fermionic chargons have been studied elsewhere \cite{SS09,DCSS15b,SSDC16}, 
and more recent work has
examined the evolution of the Hall coefficient as a function of electron density \cite{AMSY2016}. Our focus in the present paper is
on $\mathbb{Z}_2$-FL* states with low energy {\it bosonic\/} chargons. One such $\mathbb{Z}_2$-FL* state with incommensurate
spin correlations and Ising-nematic order was studied recently \cite{SCYQSSJS_2016}, 
and it exhibited a confinement transition to a
superconducting state which is usually of the Fulde-Ferrell-Larkin-Ovchinnikov type, with spatial modulation of the superconducting order.

Here, we will turn our attention to a simpler $\mathbb{Z}_2$-FL* state with bosonic chargons: this exhibits 
a direct confinement transition to a spatially uniform $d$-wave superconductor which is smoothly connected to the 
conventional BCS state. Indeed, our model for the superconducting state so obtained may be viewed as a realization of the
variational ``plain vanilla RVB" theory \cite{PlainVanilla}. A related model has been studied by Wen and collaborators \cite{Ribeiro05,Ribeiro06,MeiWen12}. We will present results on the evolution of the electronic spectrum of this
plain vanilla $\mathbb{Z}_2$-FL* state as a function of $p$, both within the superconducting and normal states.
In the superconducting state, we find that the number of gapless nodal points in the Brillouin zone (BZ) is initially twelve upon exiting the $\mathbb{Z}_2$-FL* state, but reduces to four once we are well within the confinement region. 

We also present the evolution of the Hall co-efficient in the normal state, and contrast it with the results obtained
from models of fermionic chargons. To obtain a direct transition between metallic FL* and FL states, we have to assume
a vanishing spinon pair amplitude in the FL* state at higher temperatures or fields---it is more appropriate to call this a U(1)-FL*,
although there is no formal distinction between different FL* states at non-zero temperatures.
Starting from such a U(1)-FL* state with bosonic chargons, we find,
following the results of Coleman {\it et al.} \cite{ColemanPRB2005}, a discrete jump
in the Hall coefficient from $p$ to $1+p$ at the transition from the FL* to a FL in the presence of a strong magnetic field that destroys superconductivity. No such
jump was found in the fermionic chargon approach \cite{SSDC16,AMSY2016}.

We also discuss a modification of the plain vanilla $\mathbb{Z}_2$-FL* theory to allow for translational
symmetry breaking in the confining state: this is achieved by modifying the dispersion of the bosonic chargons.
Condensing such chargons, we find a superconductor with co-existing bond density waves and pair density waves at the same 
wavevector $\K$.

We emphasize that our model is a phenomenological description of the PG phase of the underdoped cuprates, motivated by evidence of a small Fermi surface (without a broken symmetry) from transport measurements. We assume that the $\mathbb{Z}_2$-FL* with bosonic chargons is a parent state. We then show via concrete calculations that appropriate low-temperature instabilities of such a parent state can lead to $d$-wave superconductivity, as well as further density wave orders which have been observed in spectroscopic experiments. We also provide a numerical evaluation of the Hall-coefficient across a transition from a FL* to a Fermi liquid. The $\mathbb{Z}_2$-FL* phase can only arise in the presence of strong interactions between the electrons, and therefore it is quite non-trivial to establish a quantitative connection between the parameters of this phase and some more conventional model of interacting electrons, such as the $t$-$J$-$V$ model on the square lattice. As described in Ref.~\onlinecite{Punk15}, a quantitative connection between the $t$-$J$ model and a particular dimer model of the $\mathbb{Z}_2$ FL* can be made in certain limiting regimes. In the present paper we work with a more general model of the plain vanilla $\mathbb{Z}_2$-FL* state, where the parameters are fixed by demanding that the shapes of our Fermi surfaces are consistent with spectroscopic data. We hope that numerical methods like DMFT would yield accurate values of such parameters in the future.

The rest of the paper is organized as follows. In Sec.~\ref{model} we discuss and review the $\mathbb{Z}_2$-FL* 
theory with bosonic chargons,
and introduce the plain vanilla model we will focus on. In Sec.~\ref{dwave} we analyze translation symmetry preserving confinement transitions that lead to superconductivity, and the spectral function of quasiparticle excitations in the superconducting phase. 
In Sec.~\ref{HallNo} we calculate the evolution of the Hall coefficient across the confinement transition after suppressing superconductivity by a strong magnetic field. Finally, in Sec.~\ref{transBreak} we discuss the phases obtained from confinement transitions with translation symmetry breaking. We end with a discussion of the merits and demerits of the $\mathbb{Z}_2$-FL* with bosonic chargons as a candidate for the PG metal, and compare with models of the $\mathbb{Z}_2$-FL* with fermionic chargons.

\section{Model of $\mathbb{Z}_2$-FL* with bosonic chargons}
\label{model}
We begin with a brief review of the topological aspects of the $\mathbb{Z}_2$-FL*, following Ref.~\onlinecite{SCYQSSJS_2016}. For a time-reversal invariant insulating $\mathbb{Z}_2$ spin liquid, the spectrum can be described in terms of four ``superselection'' sectors, labeled as $1$, $e$, $m$ and $\epsilon$ \cite{Kitaev03}. In Schwinger boson theories of spin liquids, the $S=1/2$ bosonic spinon, carrying $\mathbb{Z}_2$ gauge charge, itself belongs to the $e$ sector. The spinless $\mathbb{Z}_2$ gauge flux, or the vison, belongs to the $m$ sector. The fused state of the bosonic spinon $e$ and the vison $m$ is the fermionic spinon or the $\epsilon$ particle, which also carries a $\mathbb{Z}_2$ gauge charge. In the metallic $\mathbb{Z}_2$-FL* state, we can augment the insulating classification by counting the charge, $Q$, of fermionic electron-like quasiparticles. To each insulating sector, we can add a spectator electron $c$, and label the resulting states as $1_c$, $e_c$, $m_c$ and $\epsilon_c$. The above discussion is summarized in Table~\ref{tab:z2}.

\begin{table}[h]
\begin{tabular}{| c || c | c | c | c || c | c | c | c ||}
\hline 
 & $1$ & $e$ & $m$ & $\epsilon$ & $1_c$ & $e_c$ & $m_c$ & $\epsilon_c$ \\
\hline \hline
$S$ & 0 & 1/2 & 0 & 1/2 & 1/2 & 0 & 1/2 & 0 \\
\hline
Statistics & boson & boson & boson & fermion & fermion & fermion & fermion & boson \\
\hline
Mutual semions & $-$ & $m$, $\epsilon$, $m_c$, $\epsilon_c$ & $e$, $\epsilon$, $e_c$, $\epsilon_c$ & $e$, $m$, $e_c$, $m_c$ &
 $-$ & $m$, $\epsilon$, $m_c$, $\epsilon_c$ & $e$, $\epsilon$, $e_c$, $\epsilon_c$ & $e$, $m$, $e_c$, $m_c$ \\
 \hline
 $Q$ & 0 & 0 & 0 & 0 & 1 & 1 & 1 & 1 \\
 \hline
 Field operator & $-$ & $b$ & $\phi$ & $f$ & $c$ & $-$ & $-$ & $B$ \\
\hline
\end{tabular}
\caption{Characteristics of sectors of the spectrum of the $\mathbb{Z}_2$-FL* state. The first four columns are the familiar sectors of an insulating spin liquid. The value of $S$ indicates integer or half-integer representations of the SU(2) spin-rotation symmetry. The ``mutual semion'' row lists the particles which have mutual seminionic statistics with the particle labeling the column. The electromagnetic charge is $Q$. The last four columns represent $Q=1$ sectors present in $\mathbb{Z}_2$-FL*, and these are obtained by adding an electron-like quasiparticle, $1_c$, to the first four sectors. The bottom row denotes the fields operators used in the present paper to annihilate/create particles in the sectors.}
\label{tab:z2}
\end{table}

Following the discussion of topological aspects, we introduce the following Hamiltonian which realizes the plain vanilla $\mathbb{Z}_2$-FL* to study the dynamics:
\beq
H = H_f + H_c + H_b 
\label{Htotal}
\eeq
$H_f$ is a mean-field Hamiltonian which describes the fermionic spinons $f$ of the $\mathbb{Z}_2$ spin liquid at $p=0$.
\beq
H_f =  - \sum_{\r \r^{\prime}, \sigma} ( \chi_{\r \r^{\prime}} + \mu_f \, \delta_{\r \r^{\prime}}) f^{\dagger}_{\r \sigma}  f_{\r^{\prime} \sigma}  +  \sum_{\r \r^{\prime}} \Delta^{f}_{\r \r^{\prime}} \epsilon_{\alpha \beta}  f^{\dagger}_{\r  \alpha}  f^{\dagger}_{\r^{\prime} \beta} + \mbox{H.c.} 
\label{Hf}
\eeq
where the chemical potential $\mu_f$ is adjusted so that $\langle f^{\dagger}_{\r \sigma}  f_{\r \sigma} \rangle = 1$ on every site, and the spinon-hopping $\chi_{\r \r^{\prime}}$ and spinon-pairing $\Delta^{f}_{\r \r^{\prime}}$ need to be determined self-consistently. 
We will take the spinon hopping to be descended directly from the electron dispersion in the cuprates, and so have no background flux.
The spinon pairing will be taken to have a $d$-wave form, as specified below.

In the FL* phase, we also require dopant charge carriers which have the same quantum number as the electron and are neutral under the internal $\mathbb{Z}_2$ gauge field. These are the $c$ fermions, which are analogous to the green dimers in the lattice model described in Ref.~\onlinecite{Punk15}. For these fermions, we choose a phenomenological dispersion $E_c(\k)$ which has hole pockets centered at $(\pm \pi/2, \pm \pi/2)$; most of the remaining discussion (barring section \ref{HallNo}) will not require the explicit nature of the dispersion:
\beq
H_c = \sum_{\k, \sigma} \xi_{\k} \,c^{\dagger}_{\k \sigma} c_{\k \sigma}, ~~~ \xi_{\k} = E_c(\k) - \mu_c
\label{Hc}
\eeq

Finally, we need to consider the coupling between the $f$ spinons and the $c$ electrons. This can be obtained from a decoupling of the Kondo coupling between the $f$ and $c$ spins via a Coqblin Shrieffer transformation \cite{CS_PRB69} appealing to a large $N$ generalization of SU(2) spins \cite{ReadNewns_JPC83}. However, here we restrict ourselves to a simple mean-field decoupling in terms of the spin singlet bosonic chargons $B_{1/2}$ (the $\epsilon_c$ particle of Table \ref{tab:z2}) with spatially local form factors $F/\tilde{F}$:
\beq
H_b &=&  \sum_{\r, \rp}  (B_{1} F_{\r \rp})^* f^{\dagger}_{\r \sigma}c_{\rp \sigma} + (B_{2}\tilde{F}_{\r \rp})^* \epsilon_{\alpha \beta} c_{\r \alpha} f_{\rp \beta} + \mbox{H.c.} \nn
B_{1} F_{\r \rp} &\sim& f^{\dagger}_{\r \sigma}c_{\rp \sigma}, ~~~ B_{2} \tilde{F}_{\r \rp} \sim \epsilon_{\alpha \beta} c_{\r \alpha} f_{\rp \beta}
\label{Hb}
\eeq

\begin{figure}
\begin{center}
\includegraphics[scale=0.8]{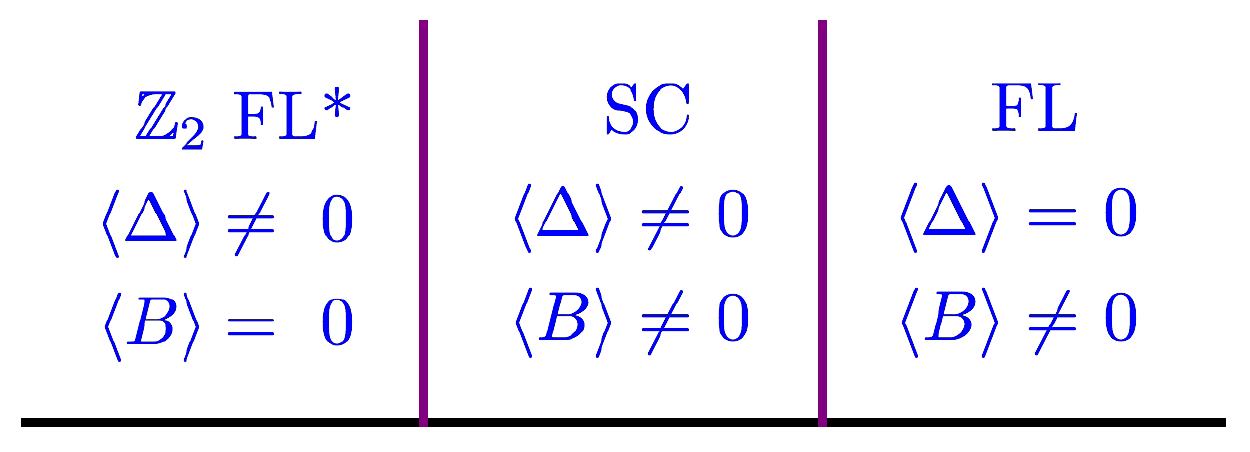}
\caption{(Color online) Schematic phase diagram of some phases arising from the $\mathbb{Z}_2$-FL*.
A FL* state with a spinon Fermi surface can be obtained at large magnetic fields by having both 
$\langle \Delta \rangle$ and $\langle B \rangle$ vanish [such a state would be a U(1)-FL* at zero $T$].}
\label{MFPhaseDia}
\end{center}
\end{figure}

Now we can discuss the phases of the Hamitonian in Eq.~(\ref{Htotal}) which are of interest to us in this paper. The $\mathbb{Z}_2$-FL* is realized when $\Delta^{f}_{\r \rp} \neq 0$, and the bosonic chargons are gapped, i.e, $\langle B_{1/2} \rangle =0$. This is the phase with Fermi pockets of the $c$ fermions. The condensation of $B_{1/2}$ (the $\epsilon_c$ particle in Table \ref{tab:z2}) leads to confinement of the $\mathbb{Z}_2$ gauge field and induces superconductivity of the electron-like $c$ fermions. At high magnetic fields, we expect a suppression of superconductivity, which leads to a FL with a large Fermi surface. This phase has $\Delta^{f}_{\r \rp} = 0$ and $\langle B_{1} \rangle \neq 0$ (but $\langle B_{2} \rangle = 0$), and the $f$ spinon acquires a charge \cite{ColemanPRB2005} and therefore contributes to charge transport together with the $c$ electron. A mean-field phase diagram is presented in Fig.~\ref{MFPhaseDia}.

One may ask whether the phases we have described are stable beyond the mean-field level, once we include the effects of fluctuations. Here, we argue that this is indeed the case at $T=0$. As long as the Kondo coupling between the $c$ and the $f$ fermions is weak, the gap to the vison (m) excitation persists, and therefore the quantum numbers of the excitations of the $\mathbb{Z}_2$ FL* state are topologically protected at $T=0$ in $d=2$ spatial dimensions \cite{FFL}. Once we are in any of the confined phases (superconductor or Fermi liquid), the appearance of the Higgs condensate $\langle B_{1/2} \rangle$ implies that the gauge fluctuations are strongly quenched. Hence, these phases are expected to be stable as well. The only point of concern is the $U(1)$ FL* which is obtained by destroying the spinon-pairing in Section \ref{HallNo}. Such a phase is known to be unstable in $d=2$ to confinement with translation symmetry breaking on the square lattice \cite{ReadSachdev}. However, if the confinement length scale is very large, the fermions should effectively realize a $U(1)$ FL* state. This picture of an effective deconfined phase has been supported by DMRG studies of a particular dimer model of the $U(1)$ FL* \cite{JSW2016}.

We end this section with a brief discussion of existing literature on confinement transitions out of a FL* phase with concomitant destruction of topological order. A superconducting transition from a specific $\mathbb{Z}_2$-FL*, corresponding to a $\mathbb{Z}_2$ spin liquid with favorable energetics and Ising-nematic order on the square lattice \cite{ReadSachdev,ReadSachdev2,SubirQPT}, was studied in Ref.~\onlinecite{SCYQSSJS_2016}. The projective transformations of the fermionic spinon $\epsilon$ under lattice symmetry operations typically led to spontaneous breaking of translation symmetry, time reversal symmetry or both. In contrast, in this paper we look for transitions to superconducting phases which arise from the plain vanilla $\mathbb{Z}_2$-FL* state described above. 
We also note that separate confinement transitions out of the $\mathbb{Z}_2$-FL* can lead to long range antiferromagnetic order \cite{CSS94,Kaul08,TGTS09} when the $e$-boson (see Table \ref{tab:z2}) condenses, or to a metallic phase with density wave order \cite{PCAS16} when the $m$-boson condenses. Quite remarkably, all three confinement transitions, obtained by condensing the bosons in Table \ref{tab:z2}, correspond to observed instabilities in the hole-doped cuprates. A detailed discussion of such unconventional metallic phases, quantum phase transitions and their relevance to the cuprate phase diagram appeared recently in Ref.~\onlinecite{SSDC16}.

\section{Confinement transitions to translation invariant superconductors}
\label{dwave}

In this section, we describe the superconducting state obtained by a confinement transition that preserves translation invariance for generic spinon-pairing. Later, we assume that the spinon-pairing form-factor $\Delta^f_{\k}$ is d-wave, and demonstrate that the resulting superconductor is also a d-wave superconductor that has spectral properties consistent with the cuprates.

\subsection{Induced superconductivity of the $c$ fermions}

The $B$ bosons carry both $\mathbb{Z}_2$ gauge charge and electromagnetic charge $e$. Therefore, their condensation is a Higgs transition that results in loss of the $\mathbb{Z}_2$ topological order. Further, the pairing of the $f$ fermions now induce a pairing between the $c$ fermions, and therefore the confined state is a superconductor. For the plain vanilla 
projective symmetry group (PSG) \cite{WenSqLattice} of the $\mathbb{Z}_2$-FL*, we can construct an effective bosonic Hamiltonian $h_{B}(\k)$ (described in detail in Ref.~\onlinecite{SCYQSSJS_2016}), and look at its dispersion. The minima of the boson dispersion would determine the wave-vector at which the B bosons condense. Here we analyze the effects of condensation of $B$ at $\Q = 0$, so that we end up with translation invariant superconductors. Note that this is allowed by the trivial PSG of the $f$ fermions in Eq.~(\ref{Hf}) for the plain vanilla $\mathbb{Z}_2$-FL* state. The discussion of translation symmetry broken superconductors is presented in Sec.~\ref{transBreak}. 

Using translation invariance to go to momentum space, the Hamiltonian in Eq.~(\ref{Htotal}) can be recast in terms of a four-component Nambu spinor $\Psi_{\k}$ as follows (neglecting a constant energy off-set):
\beq
H_{mf} = \sum_{\k} \Psi^{\dagger}_{\k} h(\k) \Psi_{\k}, \text{ where }
h(\k) = \begin{pmatrix}  \xi_{\k} & 0 & B_1 & -B_2 \\
0 & - \xi_{\k} & -B_2^{*} & -B_1^{*}  \\
B_1^{*} & -B_2 & \varepsilon_{\k} & \Delta^f_{\k} \\
-B_2^{*} & -B_1 & \Delta_{\k}^{f *} & -\varepsilon_{\k}
\end{pmatrix}
\text{ , } \Psi_{\k} = \begin{pmatrix}
c_{\k \uparrow} \\
c^{\dagger}_{-\k \downarrow}  \\
f_{\k \uparrow} \\
f^{\dagger}_{-\k \downarrow}
\end{pmatrix}
\label{H_mf}
\eeq
where we have restricted ourselves to simple on-site form factors ($F_{\r \rp}, \tilde{F}_{\r \rp} \sim \delta_{\r \rp}$). Now, we can write down the partition function in imaginary time as follows:
\beq
Z &=& \int \mathcal{D}(\bar{\Psi}, \Psi )  e^{-S}  \text{,  where } S = \sum_{\k, i \omega_n} \bar{\Psi}(\k, i \omega_n) \left[ - i \omega_n + h(\k)\right] \Psi(\k, i \omega_n)
\eeq
Since this is a Gaussian theory, we can integrate out the $f$ spinons and find an effective action for the $c$ fermions. In order to do so, we write the four-component Nambu spinor $\Psi(\k, i \omega_n)$ in terms of two two-component spinors $\psi_c$ and $\psi_f$ as follows:
\beq
\Psi(\k, i \omega_n) = \begin{pmatrix}
\psi_{c}(\k, i \omega_n) \\
\psi_{f}(\k, i \omega_n)
\end{pmatrix} \text{ where }
\psi_{c}(\k, i \omega_n) = \begin{pmatrix}
c_{\k \uparrow}(i \omega_n) \\
c^{\dagger}_{-\k \downarrow} (-i \omega_n)
\end{pmatrix}  \nonumber \\
\eeq
and $\psi_f$ is defined analogously. In terms of this the imaginary time action can be recast as [suppressing the indices $(\k, i \omega_n)$ for clarity]:
\beq
S &=& \sum_{\k, i\omega_n} \begin{pmatrix}
\bar{\psi}_c & \bar{\psi}_f
\end{pmatrix} \begin{pmatrix}
G_c^{-1} & B \\
B^{\dagger} & G_f^{-1}
\end{pmatrix} \begin{pmatrix}
\psi_c \\ \psi_f
\end{pmatrix}, \text{ where } \nn
G_c^{-1} &=& \begin{pmatrix}
- i \omega_n +  \xi_{\k} & 0 \\
0 & - i \omega_n -  \xi_{\k}
\end{pmatrix}, 
G_f^{-1} = \begin{pmatrix}
- i \omega_n +  \epsilon_{\k} & \Delta^f_{\k} \\
\Delta_{\k}^{f *} & - i \omega_n -  \epsilon_{\k}
\end{pmatrix}, 
\text{ and } B = \begin{pmatrix}
B_1 & -B_2 \\
 -B_2^{*} & -B_1^{*}
\end{pmatrix} \nn
\eeq
Now we integrate out the $f$ spinons using standard Grassman integration, resulting in the following effective action for the $c$ fermions:
\beq
S^{eff}_{c} =  \sum_{\k, i\omega_n} \bar{\psi}_c \left( G_c^{-1} - B G_f B^{\dagger} \right) \psi_c
\eeq
The diagonal elements of the second term result in self-energy corrections to the $c$ fermion pcropagator, whereas the off-diagonal elements contain information about the induced paring of the $c$ fermions. We interpret the upper off diagonal element in the effective action $S^{eff}_{c}$ as:
\beq
\sum_{\k, i \omega_n} \Delta^c(\k,i\omega_n) c^{\dagger}_{\k,\uparrow}(i\omega_n) c^{\dagger}_{-\k,\downarrow}(-i \omega_n), \text{ where }  \Delta^c(\k,i\omega_n) =  \frac{B_1^2 \Delta^f_{\k} - B_2^2 \Delta_{\k}^{f *} - 2  B_1 B_2 \epsilon_{\k}}{\epsilon_{\k}^2 + \big|\Delta^f_{\k}\big|^2 - (i \omega_n)^2} \nn
\eeq
For temperatures much smaller than the Fermi energy, we can ignore the frequency ($\omega_n$) dependence at small frequencies, since the pairing will be induced between the low-energy $c$ fermions near the Fermi surface which have finite momenta but nearly zero energy. Therefore, we set $i \omega_n = 0$ in the above expression to arrive at the main result of this section:
\beq
\Delta^c(\k) = \frac{B_1^2 \Delta^f_{\k} - B_2^2 \Delta_{\k}^{f *} - 2  B_1 B_2 \epsilon_{\k}}{\epsilon_{\k}^2 + \big|\Delta^f_{\k}\big|^2}~, \text{ as } T \rightarrow 0
\label{indSC}
\eeq
This shows that the pairing of the $f$ spinons induces a pairing of the $c$ fermions. In particular, assuming that the spinon-pairing $\Delta^{f}_{\k}$ is real, in the regimes where one condensate is much stronger than the other, i.e, $B_{1}/B_{2} \gg 1$ or $\ll 1$, we can neglect the cross-term, and the c-pairing has approximately the same form-factor as the f-pairing. For example, in the regime $B_{1}/B_{2} \gg 1$, we find that:
\beq
\Delta^c(\k) =  \frac{B_1^2 \Delta^f_{\k} }{\epsilon_{\k}^2 + \big|\Delta^f_{\k}\big|^2} 
\eeq
On the $c$ Fermi surface given by $\xi_{\k}=0$, which will generically be away from the $f$ Fermi surface given by $\epsilon_{\k} = 0$, the denominator causes a small amplitude modulation and the $c$ superconductivity will be roughly proportional to $\Delta^f_{\k}$. Therefore, in this regime, $d$-wave pairing between the spinons leads to $d$-wave pairing of the $c$ fermions as well upon condensation of the bosons. We comment that $B_{1}/B_{2} \gg 1$ is also the regime with experimentally observed spectral properties of the cuprates, as discussed in Sec.~\ref{subsec:spec}.

\subsection{Spectrum for nodal superconductivity}
\label{subsec:spec}
In this subsection, we discuss the spectrum of the $d$-wave superconductor obtained via the confinement transition, with particular focus on the number of nodal quasiparticles. In the presence of superconductivity $\Delta^{f}_{\k}$ of the $f$ fermions, we showed in the previous section that superconductivity with an identical form factor will be induced in the $c$ fermions as well, upon the confinement transition. Consider a large $f$ Fermi surface (analogous to the overdoped FL phase of the cuprates). Right after the transition, both $c$ and $f$ Fermi surfaces correspond to zero-energy quasiparticles which have charge e and spin half. Hence, if the $f$ superconductivity is $d$-wave, i.e, $\Delta^{f}_{\k} = \Delta_{d}(\text{cos}k_x - \text{cos}k_y)$, then there will be four nodal points on the $f$ Fermi surface, and eight more nodal points for the $c$ pockets as the nodal line intersects each pocket twice. We show that once one gets well into the confined phase by increasing the condensate strength $B_{1/2}$ the number of nodal points reduces to four, as observed by spectroscopic probes. Since we preserve the full $C_4$ square lattice symmetry, we restrict ourselves to studying one quarter of the full BZ ($0 \leq k_x, k_y \leq \pi$). 

\begin{figure}[h!] 
\begin{subfigure}[t]{0.45\textwidth}
\includegraphics[scale=0.5]{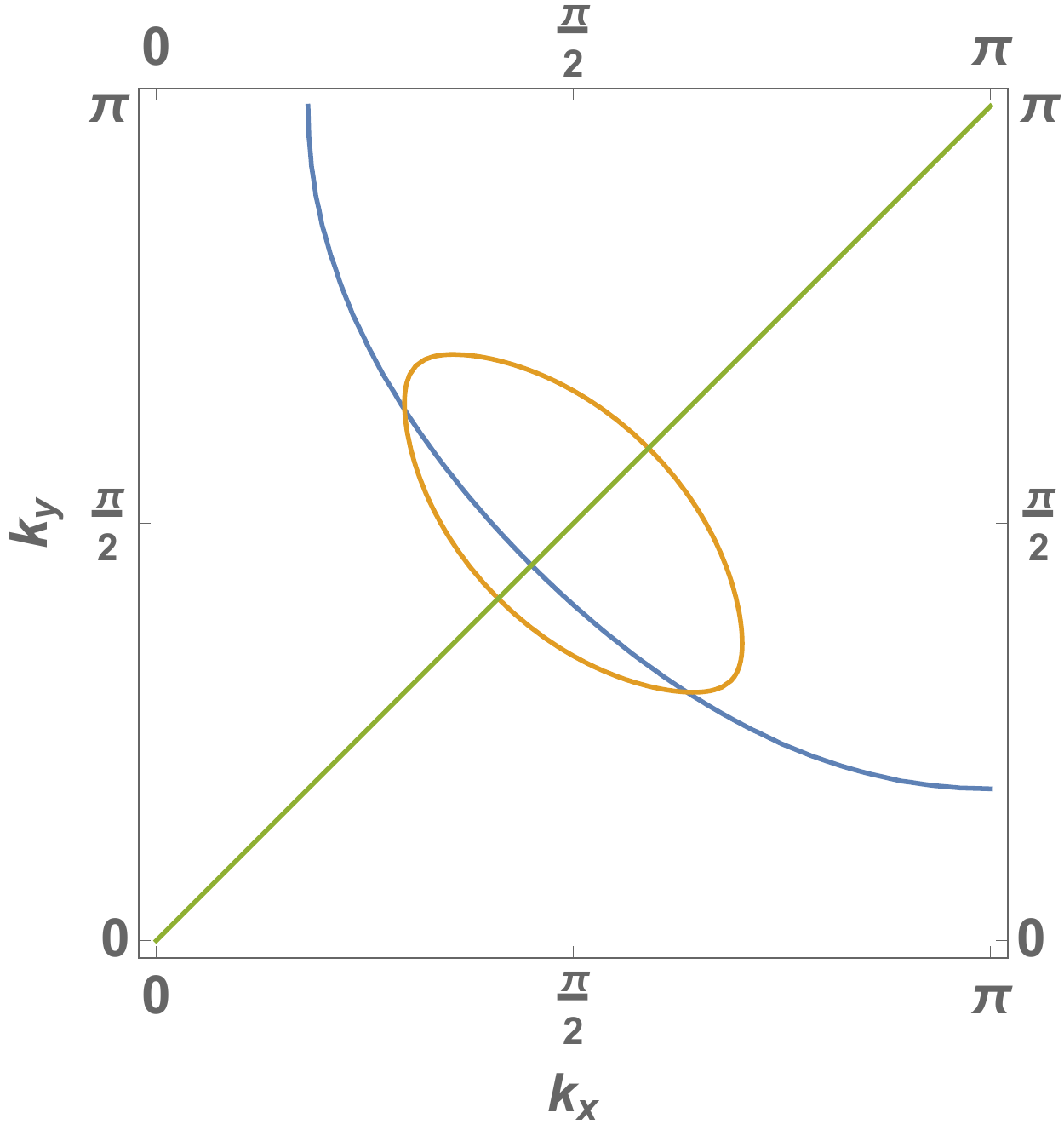} 
\caption{Fermi surface of the $f$ fermions when $\Delta^{f} = 0$ (blue), Fermi pocket of $c$ fermions (yellow) and the nodal line $k_x = k_y$ (green) in part of the full BZ}
\label{fig:subim1}
\end{subfigure}
\begin{subfigure}[t]{0.45\textwidth}
\includegraphics[scale=0.65]{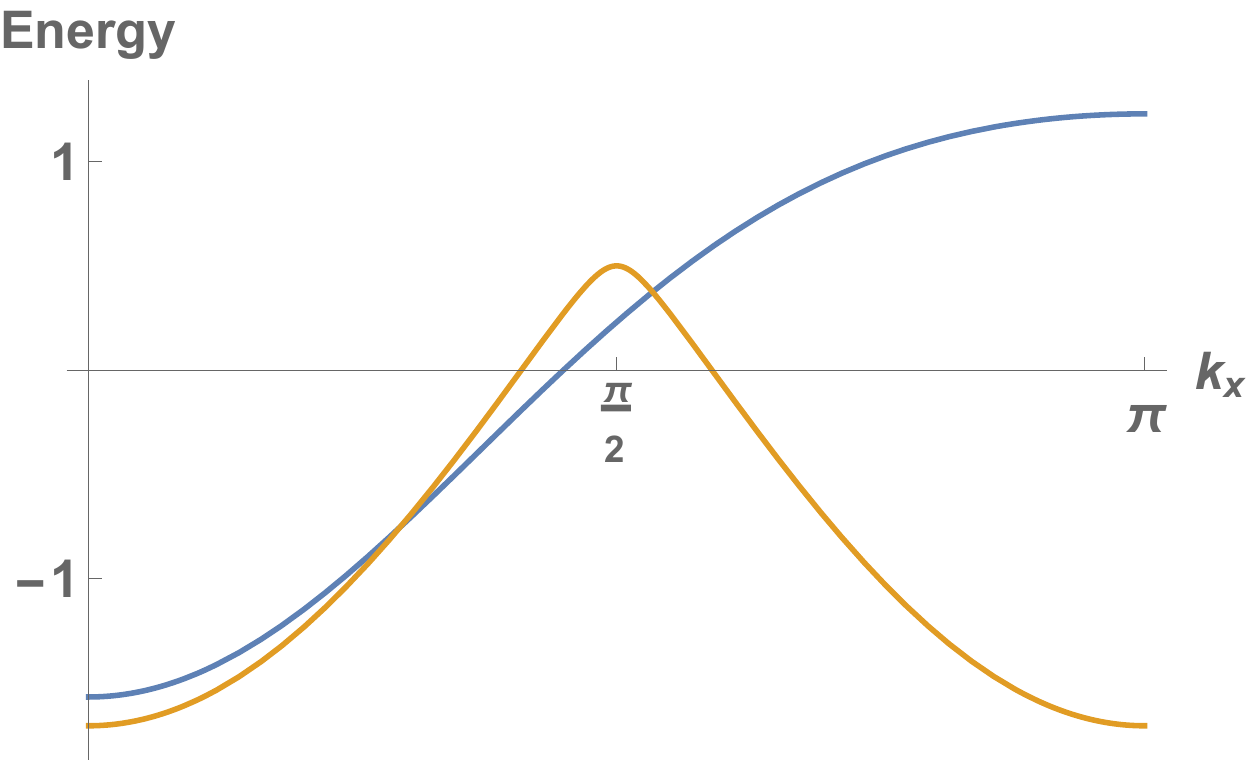}
\caption{Cuts of the dispersions $\xi_{\k}$ of the $c$ fermions (yellow) and $\epsilon_{\k}$ of the $f$ spinons (blue) plotted along the nodal line $k_x = k_y$ with $\Delta^f = 0$ } 
\label{fig:subim2}
\end{subfigure}
\caption{(Color online) Simple model of the $f$ and $c$ Fermi surfaces when $\Delta^{f} = 0$ and $B_i = 0$ }
\label{fig:SimpleModel}
\end{figure}

\begin{figure}[h!] 
\includegraphics[scale=0.8]{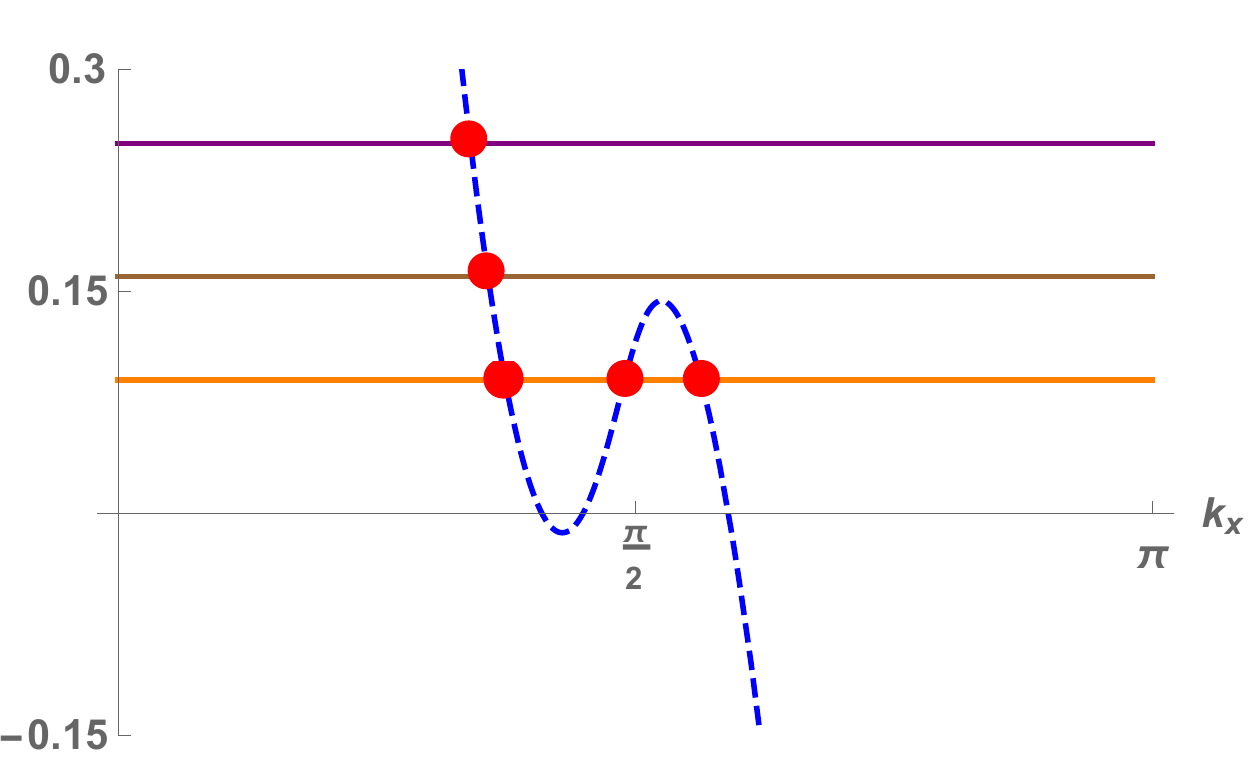}
\caption{(Color online) $\epsilon_{\k} \, \xi_{\k}  $ (dotted blue) along the nodal line $k_x = k_y$, $B_1^2 = 0.25$ (purple), $B_1^2 = 0.16$ (brown) and $B_1^2 = 0.09$ (orange). The intersection points, corresponding to the nodes, are marked with red dots. As argued in the text, there are three nodes for small $B_1$, and only one for large $B_1$ for $0 \leq k_x \leq \pi$.}
\label{Nodes_SimpleModel}
\end{figure}

We illustrate the evolution of the nodes with a generic model which has a $c$ fermion pocket centered at $\K_o = (\pi/2, \pi/2)$, coupled to the $f$ spinons, with a large Fermi surface plotted in Fig.~\ref{fig:SimpleModel}. We now use the mean field Hamiltonian in Eq.~(\ref{H_mf}) to find the nodes of the excitations when we turn on $\Delta^f_{\k}$ and condense $B_i$. The resultant excitations have a pair of doubly spin-degenerate bands, which are given by:
\beq
E_{\pm}^2 &=& \vec{B}^2 + \frac{1}{2}\left[ (\Delta^{f})^2 + \epsilon^2 + \xi^2\right] \nn 
&& \pm \frac{1}{2} \sqrt{ \left[ (\Delta^{f})^2 + \epsilon^2 - \xi^2 \right]^2 + 4\vec{B}^2\left[ (\Delta^{f})^2 + \epsilon^2 + \xi^2 \right] + 8 \epsilon \, \xi (B_1^2 - B_2^2)  + 16  B_1 B_2 \, \Delta^{f} \, \xi} ~, \nn 
\label{ExEn}
\eeq
where $\vec{B} = (B_1,B_2)$, which we have assumed to be real under appropriate gauge choice, and we have also suppressed the index $\k$ for clarity. From Eq.~(\ref{ExEn}) we can see that $E_{+}(\k) \neq 0$ whenever $\vec{B}^2 \neq 0$, so two degenerate bands are completely gapped. The condition for finding a gapless point in $E_{-}(\k)$ can be reduced to:
\beq
(B_2^2 - B_1^2 + \epsilon \, \xi)^2 + (2 B_1 B_2 - \Delta \, \xi)^2 = 0
\label{nodeEq}
\eeq
Let us investigate Eq.~(\ref{nodeEq}) when $B_1 \neq 0$ and  $B_2 = 0$. This implies that a gapless point has $ \epsilon_{\k} \, \xi_{\k} = B_1^2 > 0$, and $\Delta_{\k} \, \xi_{\k} = 0$. From the first condition, $\xi_{\k} \neq 0$, so we require $\Delta_{\k} = 0$ and therefore any gapless point must lie on the nodal line $k_x = k_y$. Now we look back at the dispersions on the nodal line given in Fig.~\ref{fig:SimpleModel}. Since $B_1^2 \neq 0$, we require that the product $\epsilon_{\k} \,\xi_{\k}\big|_{kx=ky} = B_1^2 > 0$. For small $B_1^2$, the modulations in the product near $k_x = k_y = \pi/2$ implies there are multiple solutions, as can be seen from Fig.~\ref{fig:SimpleModel}. However, for shallow $c$ pockets and a generic large $f$ Fermi surface which are required for consistency with spectroscopic experiments (see Sec.~\ref{FSevol} for further details on the Fermi surface evolution), only the solution corresponding to $k_x < \pi/2$ survives  large $B_1^2$. Therefore, we have a single node of excitations per quadrant of the BZ (ignoring spin degeneracy). Since  Eq.~(\ref{nodeEq}) are analytic in $B_1$ and $B_2$, turning on a small $B_2$ can shift the nodes away from the $k_x = k_y$ line. However, it cannot change the number of nodes. Hence for large $B_1$ and small $B_2$, we have the desired number of nodes. This is illustrated in Fig.~\ref{Nodes_SimpleModel}.

Similarly, one can also argue that for large $B_2$ and small $B_1$, we have a single nodal point per quadrant, assuming that the $c$ fermion band is quite shallow along the nodal line. In this case, for $B_1 = 0$ we require that $\epsilon_{\k} \, \xi_{\k} = - B_2^2 < 0$, which is only satisfied  for $k_x > \pi/2$ when $B_2$ is large enough. This does not change when we turn on a small $B_1$, as previously argued. However, the experimentally observed nodes in the cuprates are at $k_x < \pi/2$, so the previous scenario is more relevant for the cuprates. 

It is worthwhile to note here that even if the form factors $F_{\r \rp}$ and $\tilde{F}_{\r \rp}$ deviate from on-site interactions and we have extra momentum-dependent pre-factors in $B_{1/2}(\k)$, the number of nodes will not change unless $B_{1/2}(\k)$ go to zero near the nodal points. Therefore, this disappearance of the extra nodes is quite robust. Further, the shift of the nodes from the line $k_x = k_y$ is parametrically small if either $B_1$ or $B_2$ is small. A recent work \cite{GGCCCC2016} looked at the dimer model of FL* presented in Ref.~\onlinecite{Punk15}, and their mean-field treatment of bosonic spinons as low energy excitations of the spin liquid led to a $d$-wave superconductor (more accurately, an SC* with topological order) with eight nodes. But as our argument shows, using fermionic spinons and driving a confinement transition will ultimately lead to a $d$-wave superconductor with four nodes, as has been observed in photo-emission experiments \cite{DamascelliRMP2003,VishikShen2010}.

\begin{figure}[h!] 
\includegraphics[scale=0.75]{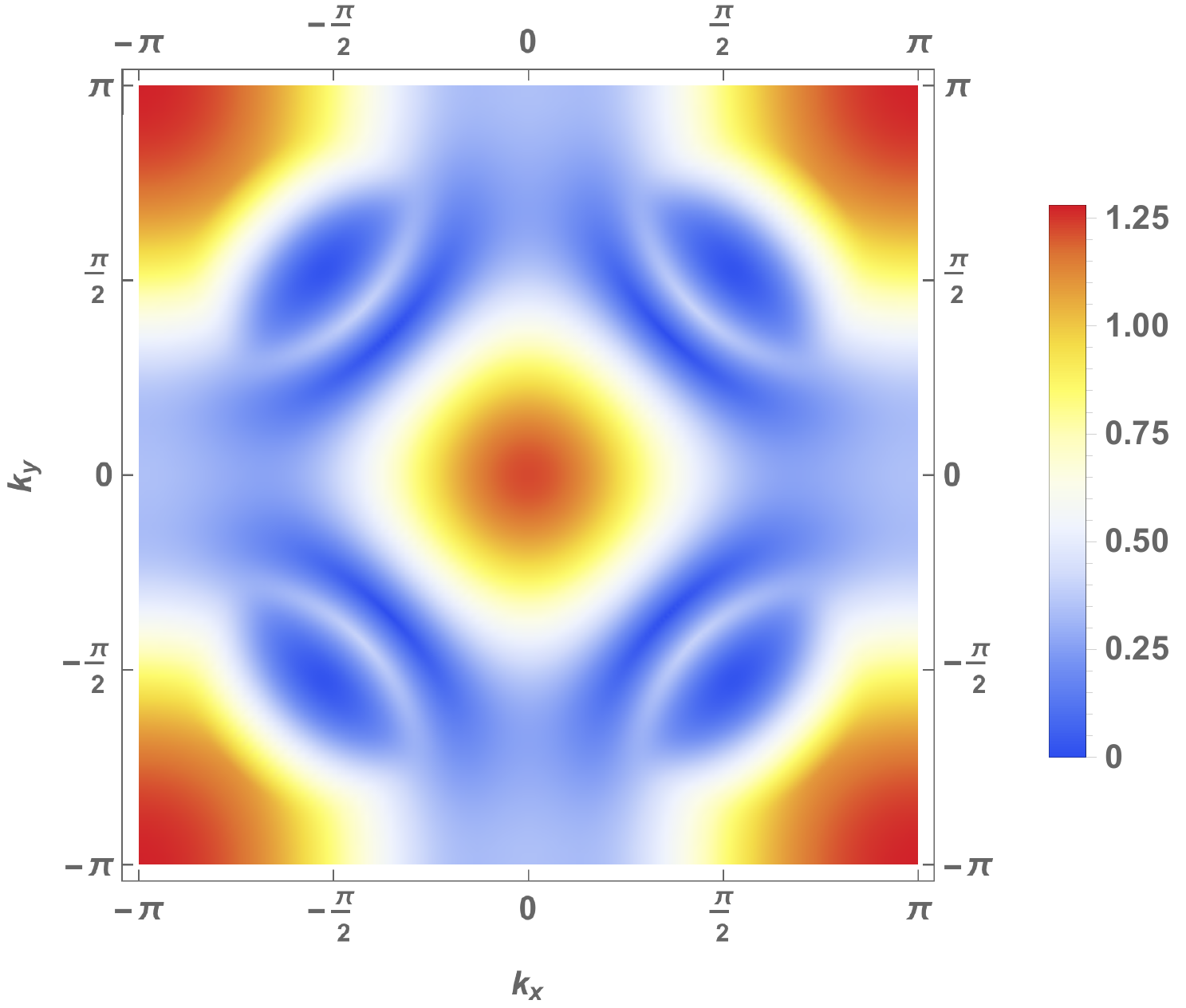}
\caption{(Color online) Schematic density plot of the energy of the lower band of quasiparticle excitations in the superconducting state, for $ \langle B_1 \rangle \neq 0$. Note the four low energy (dark blue) arcs through the nodal points.}
\label{DM_lb}
\end{figure}

Finally, we also plot the full spectrum of the Bogoliubov quasiparticles in the $d$-wave superconducting phase in this model in Fig.~\ref{DM_lb}. In the parameter regime of the superconducting phase with four nodes, the spectrum has low energy Bogoliubov excitations close to the nodes. Once we include the anisotropic quasiparticle residue $Z$ for the $c$ fermions \cite{YQSS10}, these can give rise to the Bogoliubov arcs observed in STM experiments \cite{FujitaScience2014}.

\section{Evolution of the Hall Number}
\label{HallNo}
In this section, we investigate the FL* at high magnetic fields, which suppress superconductivity.
We assume in this section that the spinon pairing $\Delta^f$ vanishes in FL* state, which makes it smoothly connected to a 
U(1)-FL* at zero $T$. 
 In the confined FL phase, self-consistency (Eq.~(\ref{indSC})) implies that neither the $c$ nor the $f$ fermions have any pairing. In absence of pairing of the $f$ spinons, their number is conserved, and therefore they couple to a  U(1) internal gauge field. However the confinement transition locks this gauge field to the external U(1) gauge field which couples to the $c$ electrons \cite{ColemanPRB2005}. Therefore, the $f$ fermions also \textit{gain a charge} and contribute to the Hall conductance. Ref.~\onlinecite{SSDC16} proposed that this can be a possible explanation for the transition of the Hall number $n_H$ (which measures the number of carriers) from $p$ to $1+p$ near optimal doping. Here we consider a concrete model and numerically evaluate $n_H$ in the FL* and FL phases to show this transition.

\subsection{Evolution from a small Fermi surface to a large one}
\label{FSevol}
In the FL* phase, the only quasiparticles carrying charge are the $c$ fermions. At the optimal doping critical point $p_c$, confinement to FL and subsequent \textit{acquirement of charge} by the $f$ fermions result in co-existence of hole-pockets with a large Fermi surface, with both quasiparticles coupling to the external electromagnetic field. As the chargon (B) condensate grows stronger, the hole pockets grow smaller and disappear, and we are left with a large Fermi surface. 

\begin{figure}[h!]
\begin{center}
\includegraphics[scale=0.6]{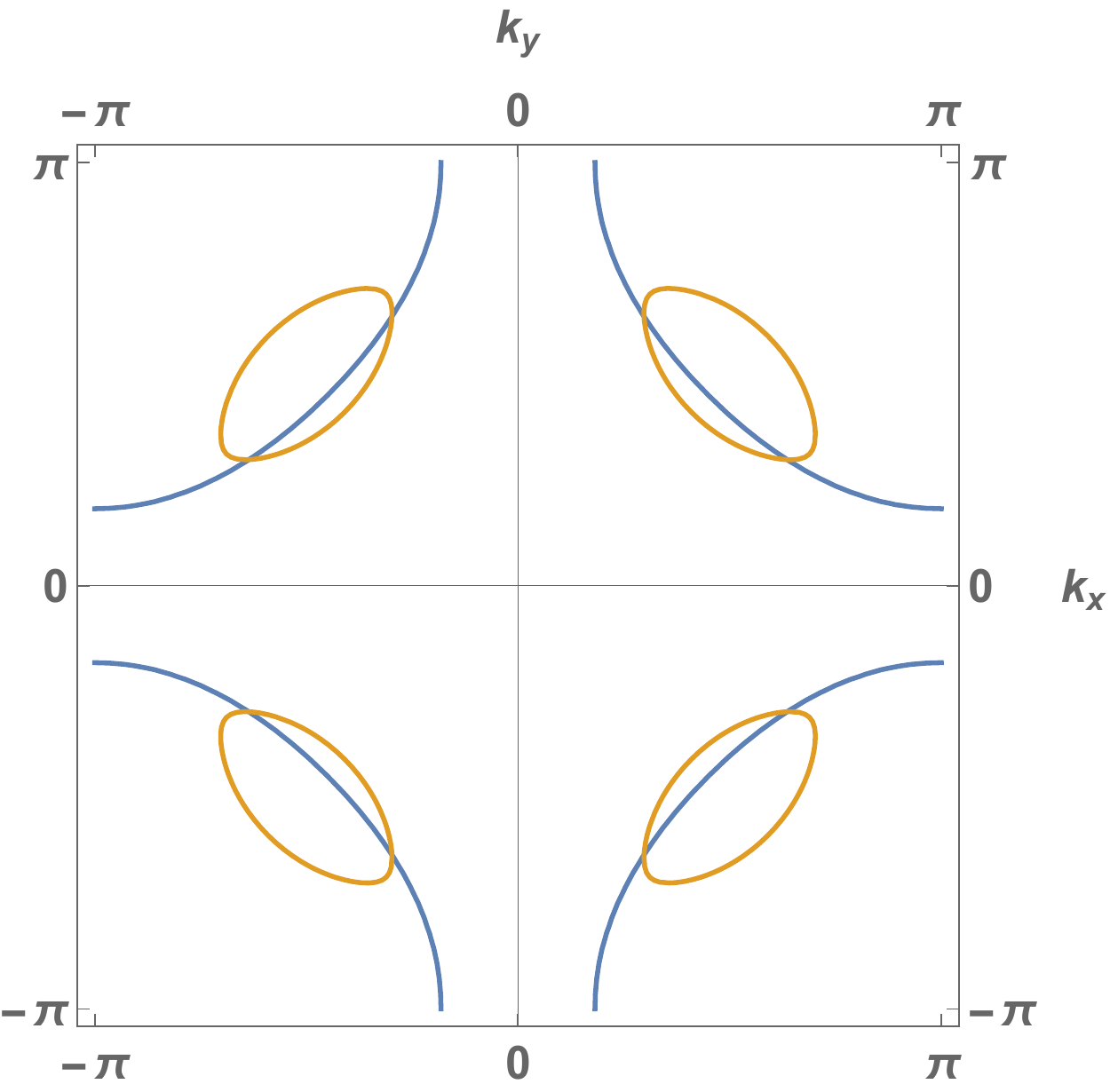}
\caption{(Color online) Plot of the Fermi surfaces from the dispersions in Eq.~(\ref{Hmf}), using $\tilde{t}_1 = 0.6,  \tilde{t}_2 = -0.2, \tilde{t}_3  = 0.1, \Delta = 0.2, \mu_c = -0.21,  t_1 = 0.35, t_2 = 0, t_3 = 0.05, \mu_f = -0.03$. Yellow curves denote the hole-like $c$ pockets. The blue contour is the large $f$ Fermi surface. }
\label{FS_B0}
\end{center}
\end{figure}

In order to write down the $c$ and $f$ band structures and their coupling $B$, we note a few desired features. Firstly, the $f$ spinons should have a large Fermi surface that ultimately resembles the generic cuprate Fermi surface in the confined phase on the overdoped side. On the other hand, the $c$ electrons should have hole-pockets around $(\pm \pi/2, \pm \pi/2)$ in the FL* phase. Such hole pockets can be obtained from a lattice model of the FL* as described in Ref.~\onlinecite{Punk15}, but we choose a slightly different phenomenological dispersion which has some additional favorable features. Just into the confined phase, both the hole-pockets and the large Fermi surface are present. We want the hole pockets to disappear for small enough values of $B$ since they would otherwise contribute extra nodes in the superconducting phase which are not observed. The large Fermi surface should not get very distorted at the doping where the hole pockets disappear, and this implies that the $c$ and $f$ Fermi surfaces have similar curvature in the overlapping region. We also want the large Fermi surface to not reconstruct into pockets or go past the van Hove filling for some range of doping after the disappearance of the hole-like Fermi pockets. All the above requirements are satisfied by the following dispersions:
\beq
H_c & = &  \sum_{\k, \sigma} \xi_{\k}   c^{\dagger}_{\k \sigma} c_{\k \sigma}, \text{ and }  H_f = \sum_{\k, \sigma} \epsilon_{\k} f^{\dagger}_{\k \sigma} f_{\k \sigma} ~~ \text{ where } \nn 
\xi_{\k} & = & - 4 \tilde{t}_2 \text{ cos}(k_x)\text{ cos}(k_y) - 2 \tilde{t}_3 \left[ \text{cos}(2k_x) + \text{ cos}(2k_y)\right] + \sqrt{ 4 \tilde{t}_1^2 \left( \text{ cos}(k_x) + \text{ cos}(k_y) \right)^2 + \Delta^2} - \mu_c, \nn 
\epsilon_{k} &=& - 2 t_1 (\text{cos}(k_x) + \text{cos}(k_y)) - 4 t_2 \text{ cos}(k_x) \text{ cos}(k_y) - 2 t_3 (\text{ cos}(2k_x) + \text{ cos}(2k_y)) - \mu_f
\label{Hmf}
\eeq
 These dispersions are plotted in Fig.~\ref{FS_B0}. 

Further, we also need the hybridization to be maximum near the pockets to suppress them quickly, and minimal at the antinodal regions to avoid significant distortion of the large Fermi surface. This can be achieved by allowing the Kondo hybridization to include further local terms in real space, beyond a simple on-site term. Moving to momentum space, we postulate a momentum-dependent form factor of the form:
\beq
B_1(\k) \equiv B_{\k} = B \left[ \text{sin}^2(k_x) + \text{sin}^2(k_y) \right]
\label{Bk}
\eeq

\begin{figure}[h!]
\begin{center}
\includegraphics[scale=0.7]{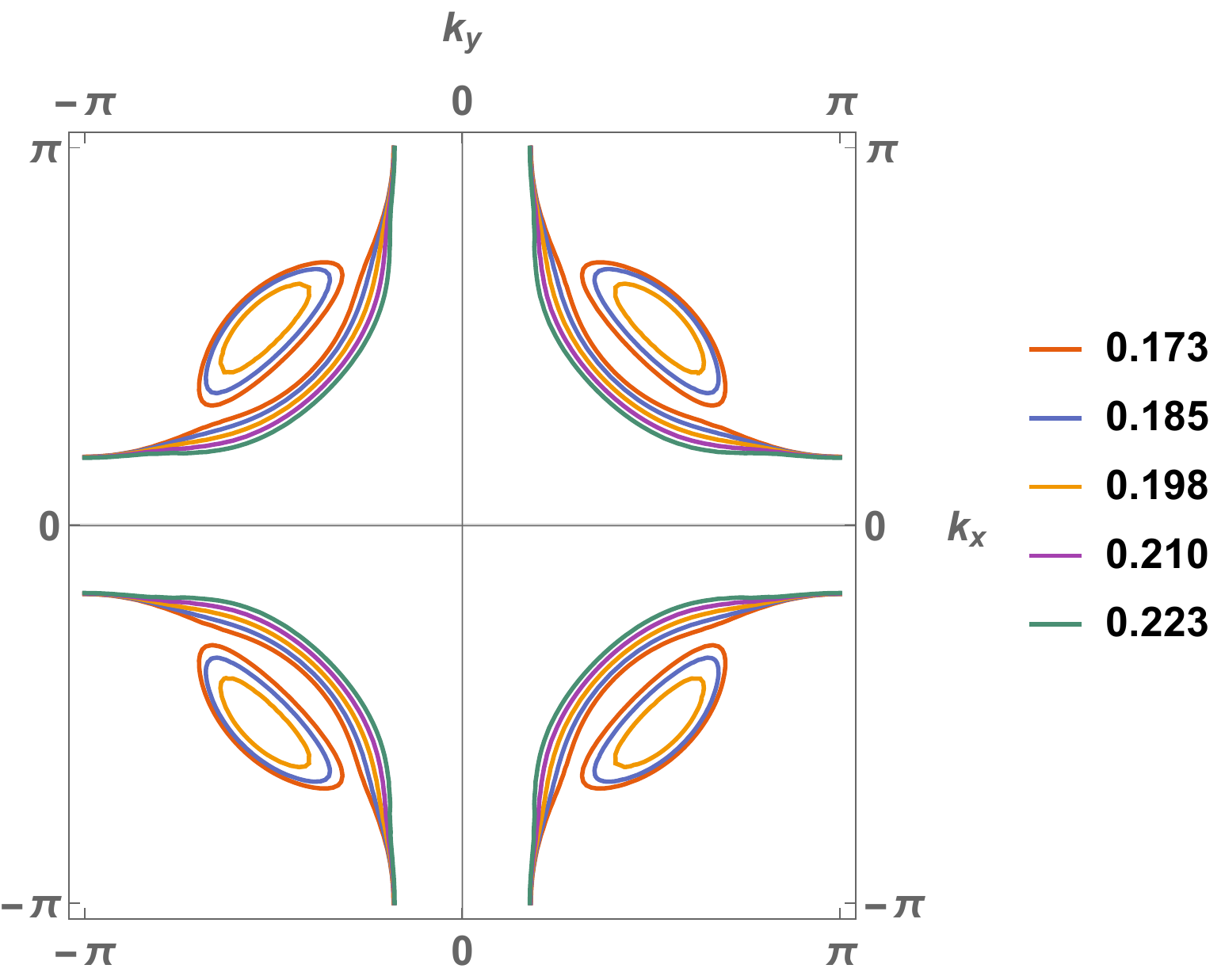}
\caption{(Color online) Evolution of the Fermi surfaces in the FL phase, color-coded by doping $p$. We chose $B = 4(p - p_c)\Theta(p-p_c)$, with $p_c  = 0.16$; hopping parameters are identical to the ones used for Fig.~\ref{FS_B0} .}
\label{bandEvol}
\end{center}
\end{figure}

Using Eqs.~(\ref{Hmf}) and (\ref{Bk}) we plot the evolution of the Fermi surfaces of the quasiparticles in the FL phase for a phenomenological $B$ that increases linearly with doping beyond the optimal doping critical point $p_c$. We first fix the chemical potentials $\mu_c$ and $\mu_f$, given the hole-doping $p$, taking into account the effect of self-energies corrections to the occupancies of $c$ and $f$ fermions in the FL phase. To do so, one can diagonalize the Hamiltonian in Eq.~(\ref{Htotal}) by the following unitary transformation in terms of new fermionic operators $ \gamma_{\k \sigma \pm}$ \cite{TSMVSS04}:
\beq
c_{\k \sigma} &=& u_{\k} \gamma_{\k \sigma +} + v_{\k} \gamma_{\k \sigma -}, ~~ f_{\k \sigma} = v_{\k} \gamma_{\k \sigma +} - v_{\k} \gamma_{\k \sigma +} ~~ \text{ where }\\
 E_{\k \pm} &=&  \frac{\epsilon_{\k} + \xi_{\k}}{2} \pm \left[ \left( \frac{\epsilon_{\k} - \xi_{\k}}{2} \right)^2 + B^2_{\k} \right]^{1/2}, ~~~~ u_{\k} = \frac{B_{\k} v_{\k}}{E_{\k +} - \xi_{\k}}, ~ u_{\k}^2 + v_{\k}^2 = 1 
 \label{2bands}
\eeq
The chemical potentials $\mu_c$ and $\mu_f$ are fixed by numerically solving the following equations:
\beq
\frac{1}{V}\sum_{\k, \sigma}  \langle c^{\dagger}_{\k \sigma} c_{\k \sigma} \rangle &=& \frac{2}{V} \sum_{\k}  u_{\k}^2 \, n_F(E_{\k +}) + v_{\k}^2 \, n_F(E_{\k -}) = 2 - p, \nn 
\frac{1}{V} \sum_{\k, \sigma}  \langle f^{\dagger}_{\k \sigma} f_{\k \sigma} \rangle &=& \frac{2}{V} \sum_{\k}  v_{\k}^2 \, n_F(E_{\k +}) + u_{\k}^2 \, n_F(E_{\k -}) = 1
\eeq
We then use these chemical potentials to calculate and plot the dispersions of the two quasiparticle bands in the FL phase in Fig.~\ref{bandEvol}. We note from Fig.~\ref{bandEvol} that the large Fermi surface remains smooth and resembles the generic cuprate Fermi surface for a range of doping beyond $p \sim 0.2$ when the hole pockets disappear.

\subsection{Calculation of $n_H$}

The charge response of the system can be calculated in the relaxation time approximation using the Boltzmann equation \cite{ziman1960EPbook} or Green's functions \cite{PVAGSJ_PRB92}. Consider a single band of charge $e$ free fermions with (grand-canonical) dispersion $E_{\k}$. In both the above approaches, the Hall conductivity of this band (for large lifetime $\tau$) can be written down as follows in terms of the velocity $v_{\alpha}(\k) = \partial_{k_{\alpha}}\varepsilon_{\k}$ and volume $V$ of the system:
\beq
\sigma_{xy}  = - \frac{e^3 \tau^2}{V} \sum_{\k, \sigma}  \left[ \frac{\partial^2 E_{\k}}{\partial k_x^2}  \frac{\partial^2 E_{\k}}{\partial k_y^2} - \left(  \frac{\partial^2 E_{\k}}{\partial k_x \partial k_y} \right)^2 \right] n_F(E_{\k})
\label{sigma_H}
\eeq
whereas the diagonal conductivity is given by:
\beq
\sigma_{\alpha \alpha} = \frac{e^2 \tau}{V}  \sum_{\k, \sigma} v_{\alpha}^2(\k)\left( - \frac{\partial n_F}{\partial E_{\k}} \right) =  \frac{e^2 \tau}{V}  \sum_{\k}  \left(  \frac{\partial^2 E_{\k}}{\partial k_{\alpha}^2} \right) n_F(E_{\k}) , \text{ where } \alpha = x,y
\label{sigma_l}
\eeq
In the FL* phase, only the $c$ fermions couple to the external gauge field and contribute to the Hall current. The Hall resistance $R_H$ can be calculated in terms of the band structure of the $c$ fermions using Eqs.~(\ref{sigma_H}) and (\ref{sigma_l}):
\beq
 R_H = \frac{\sigma_{xy}}{\sigma_{xx} \sigma_{yy}}
\eeq
In the confined FL phase, we get two bands $E_{\k \pm}$ of charge-carrying quasiparticles, with dispersion given by Eq.~(\ref{2bands}). If we neglect scattering between bands, we can just add the individual conductivity contributions of the two bands to get:
\beq
R_H = \frac{\sigma^{+}_{xy} + \sigma^{-}_{xy}}{(\sigma^{+}_{xx} + \sigma^{-}_{xx} )(\sigma^{+}_{yy} + \sigma^{-}_{yy})}
\eeq
The Hall number $n_H$, which is an approximate measure of the number of carriers, is then given by:
\beq
n_H = (R_H e)^{-1}
\eeq

\begin{figure}[h!]
\begin{center}
\includegraphics[scale=0.7]{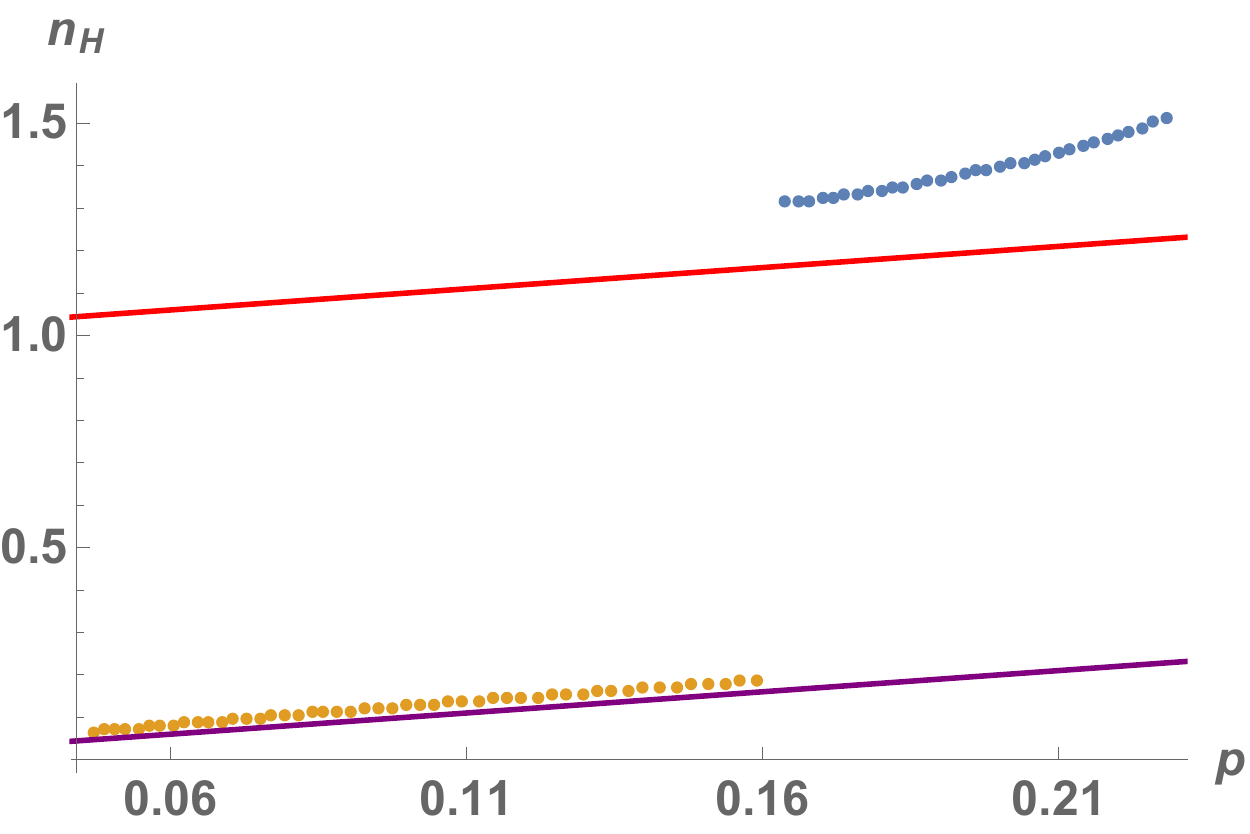}
\caption{(Color online) Numerically obtained $n_H$ as a function of doping $p$. Note the jump at $p_c = 0.16$. The purple line denotes $p$, and the red line denotes $1+p$}
\label{VaryBHallNoModFSScMu}
\end{center}
\end{figure}

We plot $n_H$ as a function of doping in Fig.~\ref{VaryBHallNoModFSScMu}, where we find that there is indeed a jump from $p$ in the underdoped regime (where we have ignored additional density wave orders) to roughly $1+p$ in the overdoped regime, at $p_c = 0.16$, where we have the transition from FL* to a Fermi liquid. The higher value of the Hall number in both phases comes from the fact that for non-circular Fermi pockets, the Hall number $n_H$ overestimates the true density of careers in the pocket (see Appendix.~\ref{app:ToyModel}). 

We note that $n_H$ changes discontinuously at the critical doping $p_c$ at $T=0$. This is generically true in a FL* with bosonic 
chargons \cite{ColemanPRB2005}, as the half-filled band of fermionic spinons will discontinuously gain a charge at the transition from FL* to FL. Upon including fluctuation corrections to the
present mean-field theory, we expect that the discontinuity will be rounded at finite temperature, but will remain a discontinuity at zero 
temperature. 

\section{Confined phases with broken translation symmetry}
\label{transBreak}
In this section, we discuss confinement transitions of the $\mathbb{Z}_2$-FL* with simultaneous breaking of translation symmetry. This happens when the bosons $B_{1/2}$ condense at finite momenta $\{ \Q_i \}$. As discussed earlier, one can determine the momenta at which this condensation occurs by a PSG analysis for a given spin liquid. However, here we restrict ourselves to a systematic analysis of the generic consequences of such a phase transition, and show that one can indeed find a phase with uniform $d$-wave superconductivity, co-existing with charge density waves $P_{\K}(\k)$ and pair-density waves $\Delta_{\K}(\k)$ at the same wave-vector $\K$, as observed in STM experiments. Well within the confined phase, the small pockets are suppressed below the Fermi level and these density waves mainly affect the large Fermi surface.

For the sake of completeness, we recall the definitions of the generalized density-wave order parameters at momenta $\{ \K_l \}$ \cite{SachdevLaPlaca,JSSS_PRB14,DCSSfeedback14}.
\beq
P_{\r \rp} = \langle c^{\dagger}_{\r \sigma} c_{\rp \sigma} \rangle = \sum_{\K_l} \left[ \frac{1}{V} \sum_{\k} e^{i \k \cdot (\r - \rp)} P_{\K_l}(\k) \right] e^{i \K_l \cdot (\r + \rp )/2} \nn 
\Delta_{\r \rp} = \langle \epsilon_{\alpha \beta} c_{\r \alpha} c_{\rp \beta} \rangle = \sum_{\K_l} \left[ \frac{1}{V} \sum_{\k} e^{i \k \cdot (\r - \rp)} \Delta_{\K_l}(\k) \right] e^{i \K_l \cdot (\r + \rp )/2}
\eeq
When $P_{\K_l}(\k)$ is independent of $\k$ ($s$-wave), then it corresponds to on-site charge density oscillations at momentum $\K_l$. When $P_{\K_l}(\k)$ is a non-trivial function of $\k$, then it corresponds to charge density oscillations on the bonds, and has also been referred to as a bond density wave in the literature \cite{SachdevLaPlaca,JSSS_PRB14,DCSSfeedback14}. In particular, we focus on a few important form factors for the bond density waves which will be relevant to this paper: 
\beq
s^{\prime} \text{ form factor with }   P_{\K_l}(\k) & \sim & \text{ cos}(k_x) + \text{ cos}(k_y), \text{ and }  \nn 
d \text{ form factor with } P_{\K_l}(\k) & \sim & \text{ cos}(k_x) - \text{ cos}(k_y)
\eeq
A similar characterization holds for the pair-density wave order parameter $\Delta_{\K}(\k)$ as well. 

We first analyze the simpler case with Ising nematic order where the fourfold rotational symmetry of the square lattice is broken to $C_2$, and then proceed to the full $C_4$ symmetric case.

\subsection{Phases in presence of nematic order}
\label{subsec:C2}
The presence of an additional Ising nematic order in the FL* state breaks the $C_4$ symmetry of the square lattice. This lifts the degeneracy between the energy eigenstates of the effective bosonic Hamiltonian $h_B$ at $\Q$ and $\hat{z} \times \Q$. However, as described in Ref.~\onlinecite{SCYQSSJS_2016}, inversion acts linearly on the $B$ bosons, so we must have $h_{B}(\Q) = h_B(-\Q)$. This implies that the condensate strengths (which are related to the components of the eigenvector of $h_{B}$) at both these momenta are the same \cite{SCYQSSJS_2016}.
Therefore, we describe the condensates as:
\beq 
\begin{pmatrix}
B_{1\r} \\ B_{2\r}
\end{pmatrix} = \begin{pmatrix}
B_{1} \\ B_{2}
\end{pmatrix} e^{i \Q\cdot \r} +  \begin{pmatrix}
B_{1} \\ B_{2}
\end{pmatrix} e^{-i \Q\cdot \r}
\label{CondC2}
\eeq
Using Eq.~(\ref{CondC2}), we can rewrite the Hamiltonian in Eq. (\ref{Hb}) as:
\beq
H_{b} = \sum_{\k} B_1^{*} ~ f^{\dagger}_{\k \sigma} c_{\k + \Q \sigma} + B_{2}^{*}\, \epsilon_{\alpha \beta} ~ c_{\k \alpha} f_{- \k - \Q \beta}  + (\Q \rightarrow -\Q) + \mbox{H.c.}
\eeq
In order to make further analytic progress, we choose $\Q$ to be commensurate, so that $N \Q$ is an integer multiple of $2\pi$ for integer $N$. Therefore, we work with the reduced BZ, where we can write the action in blocks as follows
 (suppressing the indices $ i \omega_n$ for clarity):
\beq
S =\sum_{m=0}^{N-1} \sum_{\k, i\omega_n} \begin{pmatrix}
\bar{\psi}_{c,\k + m\Q} & \bar{\psi}_{f,\k + (m+1)\Q }
\end{pmatrix} \begin{pmatrix}
G^{-1}_{c,\k + m\Q} & B \\
B^{\dagger} & G^{-1}_{f,\k +(m+1)\Q}
\end{pmatrix} \begin{pmatrix}
\psi_{c,\k + m\Q} \\ \psi_{f,\k + (m+1)\Q}
\end{pmatrix}, \text{ where }
\nn 
G^{-1}_{c,\k} = \begin{pmatrix}
- i \omega_n +  \xi_{\k} & 0 \\
0 & - i \omega_n -  \xi_{\k}
\end{pmatrix}, 
G^{-1}_{f,\k} = \begin{pmatrix}
- i \omega_n +  \epsilon_{\k} & \Delta^f_{\k} \\
\Delta_{\k}^{f *} & - i \omega_n -  \epsilon_{\k}
\end{pmatrix}, 
\text{ and } B = \begin{pmatrix}
B_1 & -B_2 \\
 -B_2^{*} & -B_1^{*}
\end{pmatrix} \nn
\eeq
Now, we can integrate out the $f$ fermions from the Gaussian action, and find an effective action for the $c$ fermions as we did in the previous section. We find that the effective action is given by:
\beq
S^{eff}_{c}  = \sum_{m=0}^{N-1} \sum_{\k, i\omega_n} && \bar{\psi}_{c,\k + m \Q} \left(G^{-1}_{c,\k + m\Q} - B \left[ G_{f,\k + (m+1)\Q} + G_{f,\k + (m-1)\Q} \right] B^{\dagger} \right) \psi_{c,\k + m\Q}  \nn 
&& +  \bar{\psi}_{c,\k + (m-1)\Q} \left( B G_{f, \k + m\Q} B^{\dagger} \right) \psi_{c, \k + (m+1)\Q} + \bar{\psi}_{c,\k + (m+1)\Q} \left( B G_{f, \k + m\Q} B^{\dagger} \right) \psi_{c, \k + (m-1)\Q} \nn 
\eeq
The diagonal terms correspond to uniform superconductivity, whereas the off-diagonal terms are responsible for density waves. Below, we describe each of these order parameters.

First, we look at uniform superconductivity. The $c$ fermion pairing term in the low frequency limit ($i \omega_n \rightarrow 0$) is given by:
\beq
\Delta^c(\k + m \Q, i \omega_n \rightarrow 0) &=& (B_1^2 - B_2^2) \left( \frac{\Delta^f_{\k + (m-1)\Q}}{\epsilon_{\k + (m-1)\Q}^2 + \big|\Delta^f_{\k + (m-1)\Q}\big|^2} + \frac{\Delta^f_{\k + (m+1)\Q}}{\epsilon_{\k + (m+1)\Q}^2 + \big|\Delta^f_{\k + (m+1)\Q}\big|^2} \right) \nn 
&  & - 2 B_1 B_2 \left( \frac{\epsilon^f_{\k + (m-1)\Q}}{\epsilon_{\k + (m-1)\Q}^2 + \big|\Delta^f_{\k + (m-1)\Q}\big|^2} + \frac{\epsilon^f_{\k + (m+1)\Q}}{\epsilon_{\k + (m+1)\Q}^2 + \big|\Delta^f_{\k + (m+1)\Q}\big|^2} \right) \nn
\eeq
If we let $\k$ belong to the full B.Z., we can re-write this term as:
\beq
\Delta^c(\k, i \omega_n \rightarrow 0) &=& (B_1^2 - B_2^2) \left( \frac{\Delta^f_{\k  - \Q}}{\epsilon_{\k}^2 + \big|\Delta^f_{\k -\Q}\big|^2} + \frac{\Delta^f_{\k + \Q}}{\epsilon_{\k}^2 + \big|\Delta^f_{\k + \Q}\big|^2} \right) -  2 B_1 B_2  \left( \frac{\epsilon^f_{\k  - \Q}}{\epsilon_{\k}^2 + \big|\Delta^f_{\k -\Q}\big|^2} + \frac{\epsilon^f_{\k + \Q}}{\epsilon_{\k}^2 + \big|\Delta^f_{\k + \Q}\big|^2} \right) \nn
\eeq
We again assume that the $f$ Fermi surface, given by $\epsilon_{\k} = 0$, is far from the $c$ Fermi surface, and take the limit of $B_1/B_2 \gg 1$. Then, we can approximately estimate the form-factor of the induced superconducting order parameter as:
\beq
\Delta^c(\k, i \omega_n \rightarrow 0) \sim \Delta^f_{\k  - \Q} + \Delta^f_{\k  + \Q} = 2 \Delta_d \left( \text{cos }Q_x \text{ cos }k_x - \text{cos }Q_y \text{ cos }k_y \right)
\eeq
This is a nodal superconductor, but the nodes are shifted from the diagonal $k_x = \pm k_y$ lines unless the ordering wave-vector $\Q$ is diagonal. However, in STM experiments \cite{hamidian_2015nphys,Hamidian16}, the observed ordering wave-vector is mainly axial --- this would result in nodes away from the diagonal lines, which is inconsistent with spectroscopic data. In Sec.~\ref{subsec:C4}, we show that in the presence of full $C_4$ symmetry, this feature goes away and we can find a $d$-wave superconductor with nodes along $k_x = \pm k_y$.

In addition to uniform $d$-wave superconductivity we find that we also have Cooper pairing at finite momentum $\K = 2\Q$, as the action $S^{eff}_{c} $ explicitly contains off-diagonal terms which are pairing between the $c$ fermions at momenta $\k + \Q $ and $-\k + \Q$. Analogous to the uniform superconducting case, we look for the following term in $S^{eff}_{c} $ to find the pair-density wave (PDW) order parameter:
\beq
\sum_{\k, i \omega_n} \Delta_{2\Q}(\k,i \omega_n) c^{\dagger}_{\k - \Q, \uparrow}(i \omega_n) c^{\dagger}_{-\k - \Q, \downarrow}(-i \omega_n) 
\eeq
Such a term is indeed present, and the PDW order parameter at low frequency ($i \omega_n \rightarrow 0$) is given by:
\beq
\Delta^c_{2\Q}(\k, i \omega_n \rightarrow 0) =  \frac{ (B_1^2 - B_2^2) \Delta^f_{\k} - 2 B_1 B_2\epsilon_{\k}}{\epsilon_{\k}^2 + \big|\Delta^f_{\k}\big|^2} 
\eeq
An analogous PDW term is present at momentum $\K =-2\Q$ as well. Again, in the limit where $B_1 \gg B_2$, we have a form-factor which is proportional to $\Delta^{f}_{\k}$. Therefore, in this regime, we have a PDW with a $d$-wave form factor.

Further, there is a charge density wave as the same wave-vector $\K = 2\Q$, as the off-diagonal term in $S^{eff}_{c} $ again breaks translation symmetry explicitly in the particle-hole channel as well. From a term in $S^{eff}_{c} $ of the form:
\beq
\sum_{\k, i \omega_n} P_{2\Q}(\k,i \omega_n) c^{\dagger}_{\k -\Q, \sigma}(i \omega_n) c_{\k + \Q, \sigma}(i \omega_n) 
\eeq
we find that the density wave order parameter is given in the low frequency limit by:
\beq
P_{2\Q}(\k, i \omega_n \rightarrow 0) = \frac{(|B_1|^2 - |B_2|^2)\epsilon_{\k} -(B_1^*B_2 + B_2^*B_1)\Delta^{f}_{\k } }{\epsilon_{\k}^2 + \big|\Delta^f_{\k }\big|^2}
\eeq
In particular, in the regime $B_1/B_2 \gg 1$, the term proportional to the $s\,(s^{\prime})$-wave $\epsilon_{\k}$ dominates. That $\epsilon_{\k}$ is predominantly s$^{\prime}$ follows from the dispersion of the $f$ spinons which must give rise to a $C_4$ symmetric large Fermi surface in the overdoped regime (as shown in Fig.~\ref{FS_B0}), to be consistent with the Fermi liquid at large doping. Therefore, this is a state with a $s\,(s^{\prime})$-wave charge density coexisting with $d$-wave superconductivity and $d$-wave PDW, provided that the $f$ superconductor has a $d$-wave form factor $\Delta^f_{\k}$. 

\subsection{Phases with full $C_4$ rotation symmetry}
\label{subsec:C4}
In the presence of full $C_4$ rotation symmetry, the simplest situation corresponds to the $B$ bosons transforming linearly (and not projectively) under $\pi/2$ rotations. This follows from the trivial PSG of the $f$ fermions in the plain vanilla $\mathbb{Z}_2$-FL*. In this case, the boson dispersion will have four minima at $\pm \Q$ and $\pm \tQ $, where $\tQ = \hat{z} \times \Q$. In the absence of extra projective phase factors, we have $h_{B}(\Q) = h_B(-\Q) = h_B(\tQ) = h_B(-\tQ)$. Therefore, we can describe the condensate as:
\beq 
\begin{pmatrix}
B_{1\r} \\ B_{2\r}
\end{pmatrix} = \begin{pmatrix}
B_{1} \\ B_{2}
\end{pmatrix} \left( e^{i \Q\cdot \r} + e^{-i \Q\cdot \r} +  e^{i \tQ\cdot \r} +  e^{-i \tQ\cdot \r} \right)
\eeq
Then, we can rewrite the Hamiltonian in Eq. (\ref{Hb}) as:
\beq
H_{b} = \sum_{\k} \left[ B_1^{*} ~ f^{\dagger}_{\k \sigma} c_{\k + \Q \sigma} + B_{2}^{*}\, \epsilon_{\alpha \beta} ~ c_{\k \alpha} f_{- \k - \Q \beta}  + (\Q \rightarrow -\Q)  + (\Q \rightarrow \tQ) + (\Q \rightarrow -\tQ) \right]+ \mbox{H.c.} \nn
\eeq
The rest of the calculations are analogous to Sec.~\ref{subsec:C2}. In the reduced BZ, the action is given by [using $(\k, m, n)$ to denote momentum $\k + m \Q + n \tQ$]:
\beq 
S =\sum_{m,n=0}^{N-1} \sum_{\k, i\omega_n}   \begin{pmatrix}
\bar{\psi}_{c,\k,m,n} & \bar{\psi}_{f,\k,m+1,n} &  \bar{\psi}_{f,\k,m,n+1}
\end{pmatrix} \begin{pmatrix}
G^{-1}_{c,\k,m} & B & B \\
B^{\dagger} & G^{-1}_{f,\k,m+1,n} & 0 \\
B^{\dagger} & 0 & G^{-1}_{f,\k,m,n+1}
\end{pmatrix} \begin{pmatrix}
\psi_{c,\k,m,n} \\ \psi_{f,\k,m+1,n} \\ \psi_{f,\k,m,n+1}
\end{pmatrix} \nn
\eeq
On integrating out the $f$ fermions, we obtain the effective action:
\beq
S^{eff}_{c}  = \sum_{m=0}^{N-1} \sum_{\k, i\omega_n} &&  \bar{\psi}_{c,\k,m,n} \left(G^{-1}_{c,\k,m,n} - B \left[ G_{f,\k,m+1,n} + G_{f,\k, m-1,n} + G_{f,\k,m,n+1} + G_{f,\k, m,n-1} \right] B^{\dagger} \right) \psi_{c,\k,m,n}  \nn 
&& +  \bar{\psi}_{c,\k,m-1,n} \left( B G_{f, \k,m,n} B^{\dagger} \right) \psi_{c, \k, m+1,n} + \bar{\psi}_{c,\k,m+1,n} \left( B G_{f, \k,m,n} B^{\dagger} \right) \psi_{c, \k,m-1,n}   \nn 
&& + \bar{\psi}_{c,\k,m,n-1} \left( B G_{f, \k,m,n} B^{\dagger} \right) \psi_{c, \k, m,n+1} + \bar{\psi}_{c,\k,m,n+1} \left( B G_{f, \k,m,n} B^{\dagger} \right) \psi_{c, \k,m,n-1}  \nn
\eeq
From this, we can deduce the form factor of the induced $c$ superconductivity, in the $i \omega_n \rightarrow 0$ and $B_1/B_2 \gg 1$ limit when $\Delta^{f}_{\k}$ is $d$-wave:
\beq
\Delta^c(\k, i \omega_n \rightarrow 0) & \sim & \Delta^f_{\k  - \Q} + \Delta^f_{\k  + \Q} + \Delta^f_{\k  - \tQ} + \Delta^f_{\k  + \tQ}  \nn
& = & 2\Delta_d (\text{cos }Q_x +  \text{cos }Q_y) \left(  \text{cos }k_x -  \text{ cos }k_y \right)
\eeq
Therefore, we have induced $d$-wave superconductivity of the $c$ fermions with nodes along the diagonal lines $k_x = \pm k_y$. This $d$-wave form factor is independent of the wave-vector $\Q$, and therefore here we can allow $\Q$ to be axial unlike the nematic case. 

A calculation similar to Sec.~\ref{subsec:C2} shows one also has co-existing charge density waves and PDW at the same wave-vector $\K = 2 \Q$ (and also at momenta related by $\pi/2$ rotations). Their descriptions are identical to the nematic case. In particular, in the $i \omega_n \rightarrow 0$ and $B_1/B_2 \gg 1$ limit, the form factors are predominantly $s\,(s^{\prime})$ for the charge density wave and $d$ for the PDW. Therefore, although the appearance of induced density waves in the particle-particle and particle-hole channels at the same-wave vector are concommitant with uniform $d$-wave superconductivity, their form factors are flipped --- in STM experiments the charge density wave has a $d$-form factor and the PDW has an $s\,(s^{\prime})$ form factor \cite{Hamidian16}.

\section{Conclusion}
We have analyzed several aspects of a plain vanilla $\mathbb{Z}_2$-FL* metal
as a possible candidate for the pseudogap phase of the high $T_c$ cuprates (a related model was studied in Refs.~\onlinecite{Ribeiro05,Ribeiro06,MeiWen12}). In particular, we have shown how to obtain a $d$-wave superconductor with consistent spectral properties via a confinement transition. This $d$-wave superconductor is very similar to that obtained
in the plain vanilla RVB theory \cite{PlainVanilla}.

We also analyzed confinement transitions accompanied by spontaneous translation symmetry-breaking, and found that a state with  charge density waves and pair density waves at the same wave-vector $\K$, together with uniform $d$-wave superconductivity falls out remarkably out of a confinement transition. A very similar state has been observed in recent STM experiments \cite{Hamidian16}, only differing in the form-factors of the associated density-wave orders. 

Finally, we also calculated the evolution of the Hall coefficient in the normal state, and demonstrated the jump from $p$ to $1+p$ across the critical point near optimal doping. We argued that at $T=0$, a FL* with bosonic chargons will always give a discontinuous jump at the transition, even after accounting for fluctuations in the mean field theory \cite{ColemanPRB2005}. The jump will be rounded by fluctuations at $T>0$, though.
The data of Refs.~\onlinecite{LTCP15,Laliberte2016} show a smooth evolution which does not sharpen upon lowering the
temperature, and so appears to be incompatible with the present model. Nevertheless, we need measurements at low $T$ to definitively 
rule out the present model of bosonic chargons. 

Other theories can better match the evolution of the Hall effect in Refs.~\onlinecite{LTCP15,Laliberte2016}. 
The simplest of these assumes the presence of the $(\pi, \pi)$ antiferromagnetic order which vanishes at
the quantum critical doping \cite{Storey_EPL}. Spiral antiferromagnetic order also yields a similar evolution \cite{AMSY2016}.
While magnetic order is clearly not present near optimal doping in zero field, the possibility of
magnetic field induced antiferromagnetism has not yet been ruled out. NMR or muon spin resonance experiments are
promising routes to settling this issue.

However, these models suggest an attractive option in which the magnetic order is not long-ranged,
but quantum-fluctuating with only intermediate range correlations. Such a model of quantum fluctuating
order leads to a $\mathbb{Z}_2$-FL* model of the pseudogap with low energy fermionic chargons \cite{SS09,YQSS10,SSDC16}.
Indeed, the evolution of the Hall effect in one of such models \cite{SSDC16} is essentially identical to that in the theory
which assumes incommensurate spiral order \cite{AMSY2016}. So the current status is that a pseudogap model
with fermionic chargons is in better accord with recent Hall effect observations \cite{LTCP15,Laliberte2016} near optimal doping
than the plain vanilla $\mathbb{Z}_2$-FL* model with bosonic chargons considered in the present paper.

\subsection*{Acknowledgements}
We thank A. Eberlein, D. Chowdhury, and A. A. Patel for valuable discussions. This research was supported by the NSF under Grant DMR-1360789.
Research at Perimeter Institute is supported by the Government of Canada through Industry Canada and by the Province of Ontario 
through the Ministry of Research and Innovation. SS also acknowledges support from Cenovus Energy at Perimeter Institute.

\appendix
\section{Overestimation of carrier densities by $n_H$ for an elliptical pocket}
\label{app:ToyModel}
We illustrate with a toy model that $n_H$ is overestimated for an elliptical pocket unless the axes of the ellipse are exactly aligned along the measurement (x-y) axes.  Consider the following elliptical pocket with axes inclined at an angle $\alpha$ to the x-y axes, such that the dispersion $\varepsilon_{\k}$ is given by:
\beq
\varepsilon_{\k} &=& \frac{(k_x \text{ cos}(\alpha) - k_y \text{ sin}(\alpha))^2}{2 m_1} +  \frac{( k_x \text{ sin}(\alpha) + k_y \text{ cos}(\alpha))^2}{2 m_2} \nn \\
& = & k_x^2 \left( \frac{\text{ cos}^2(\alpha)}{2 m_1} + \frac{\text{ sin}^2(\alpha)}{2 m_2} \right) + k_y^2 \left( \frac{\text{ sin}^2(\alpha)}{2 m_1} + \frac{\text{ cos}^2(\alpha)}{2 m_2} \right) +  k_x k_y \text{ sin}(2\alpha)\left( -\frac{1}{2 m_1} + \frac{1}{2 m_2} \right). \nn \\
\eeq
We can now use Eqs.~(\ref{sigma_H}) and (\ref{sigma_l}) to evaluate the conductivities exactly. Denoting by $n_c$ the total number of carriers in the pocket (including spin degeneracy), we find that:
\beq
\sigma_{xx} & = & e^2 \tau \left( \frac{\text{ cos}^2(\alpha)}{m_1} + \frac{\text{ sin}^2(\alpha)}{ m_2} \right)\left( \sum_{\k, \sigma} n_F(\varepsilon_{\k}) \right) = e^2 \tau \left( \frac{\text{ cos}^2(\alpha)}{m_1} + \frac{\text{ sin}^2(\alpha)}{ m_2} \right) n_c \nn \\
\sigma_{yy} & = &  e^2 \tau \left( \frac{\text{ sin}^2(\alpha)}{m_1} + \frac{\text{ cos}^2(\alpha)}{m_2} \right) n_c \nn 
\sigma_{xy}  & = & -e^3 \tau^2 \left[ \left( \frac{\text{ cos}^2(\alpha)}{m_1} + \frac{\text{ sin}^2(\alpha)}{ m_2} \right)  \left( \frac{\text{ sin}^2(\alpha)}{m_1} + \frac{\text{ cos}^2(\alpha)}{m_2} \right) - \text{ sin}^2(2\alpha)\left( -\frac{1}{2 m_1} + \frac{1}{2 m_2} \right)^2 \right] n_c \nn
\eeq
Therefore, the Hall resistance $R_H$ is given by
\beq
R_H = \frac{\sigma_{xy}}{\sigma_{xx} \sigma_{yy}} &=&  - \frac{1}{n_c e} \left[ 1 - \frac{ \text{ sin}^2(2\alpha)\left( -\frac{1}{2 m_1} + \frac{1}{2 m_2} \right)^2 }{\left( \frac{\text{ cos}^2(\alpha)}{m_1} + \frac{\text{ sin}^2(\alpha)}{ m_2} \right)\left( \frac{\text{ sin}^2(\alpha)}{m_1} + \frac{\text{ cos}^2(\alpha)}{m_2} \right)} \right] \nn 
& = &  - \frac{1}{n_c e}  \left[  \frac{1}{1 + \left( \mu - \frac{1}{\mu} \right)^2 \text{ sin}^2(\alpha) \text{ cos}^2(\alpha)}  \right] , \text{ where } \mu = \frac{m_1}{m_2}
\eeq
Therefore, $|e R_H|^{-1} =  n_c$ only if $\mu = 1$, i.e, $m_1 = m_2$, which corresponds to a circular pocket, or if $\alpha = 0 \text{ or } \pi/2$, which corresponds to having elliptical pockets with axes aligned along the measurement (x-y) axes.  Otherwise $|R_{H}|$ is smaller, implying that $|n_H| = |e R_{H}|^{-1}$ is larger than $n_c$, the true carrier density. An additional feature to note is that the deviation is larger the more anisotropic the pocket is, as well as if the angle subtended by the axes of the ellipse to the measurement axes is around $\pi/4$. Both of these features are present in our Fermi pockets, which push up the value of $n_H$ we see in the numerics.

\bibliographystyle{apsrev4-1_custom}
\bibliography{FLStarBosonicChargons}

\end{document}